\documentclass{webofc}
\usepackage[varg]{txfonts}   
\usepackage{lineno}
\usepackage{float}
\newcommand{\etaphi}{\ensuremath{(\eta,\phi)}~}

\newcommand{\ptJet}{\ensuremath{p_{\mathrm{T, jet}}}}

\newcommand{\ptJetGeq}[1]{\ensuremath{p_{\mathrm{T, jet}} \geq #1~\mathrm{GeV}/c}}

\newcommand{\ptTrack}{\ensuremath{p_{\mathrm{T, track}}}}

\newcommand{\pp}{\ensuremath{p + p}~}

\newcommand{\g}{$g$}

\newcommand{\lesub}{\ensuremath{\rm LeSub}}

\newcommand{\Gevc}{\ensuremath{~\mathrm{GeV}/c}}

\newcommand{\EtTower}{\ensuremath{E_{\mathrm{T, tower}}}}

\newcommand{\sGevAuAu}{\ensuremath{\sqrt{s_{_{\rm NN}}} = 200\ \mathrm{GeV}}}

\newcommand{\sGevPbPb}{\ensuremath{\sqrt{s_{_{\rm NN}}} = 2.76\ \mathrm{TeV}}}

\newcommand{\ptd}{\ensuremath{p_{\mathrm{T}}^\mathrm{D}}}

\newcommand{\addOnePanelInFig}[2]{\includegraphics[width = #2\linewidth]{#1}}

\begin{document}
\title{Generalized angularities measurements from STAR \\ at $\sqrt{s_{\rm NN}} = $ 200 GeV}
%
%

\author{\firstname{Tanmay} \lastname{Pani}\inst{1}\thanks{This work is supported by the National Science Foundation under Grant number: 1913624.} for the STAR Collaboration}

\institute{
	Rutgers University,\\
    136 Frelinghuysen Road, Piscataway, USA \\
     \email{tp543@physics.rutgers.edu}
}


\abstract{%
  Jets are produced in early stages of heavy-ion collisions and undergo modified showering in the quark-gluon plasma (QGP) medium relative to a vacuum case. These modifications can be measured using observables like jet momentum profile and generalized angularities to study the details of jet-medium interactions. Jet momentum profile ($\rho(r)$) encodes radially differential information about jet broadening and has shown migration of charged energy towards the jet periphery in Pb+Pb collisions at the LHC. Measurements of generalized angularities (girth $g$ and momentum dispersion $p_T^D$) and LeSub (difference between leading and subleading constituents) from Pb+Pb collisions at the LHC show harder, or more quark-like jet fragmentation, in the presence of the medium. Measuring these distributions in heavy-ion collisions at RHIC will help us further characterize the jet-medium interactions in a phase-space region complimentary to that of the LHC. In this contribution, we present the first measurements of fully corrected $g$, $p_T^D$ and LeSub observables using hard-core jets in Au+Au collisions at $\sqrt{s_{\rm NN}}=200$ GeV, collected by the STAR experiment at RHIC. 
}
\maketitle

\section{Introduction}\label{sect:Intro}

In heavy-ion collisions, collimated sprays of hadrons called jets, which are the experimental proxies for hard scattered partons from early stages, traverse the quark-gluon plasma (QGP) medium and are modified relative to a  \pp baseline. This is known as \textit{jet quenching}~\cite{Gyulassy_2004}. Therefore, jets are used as probes of the QGP to access information about the interaction between hard partons and QGP. In particular, one can study intra-jet angular distribution of energy relative to the jet-axis through generalized jet angularities, calculated as \cite{Larkoski_2014}:

\begin{equation}\label{eq:GenAngIntro}
	\lambda_\beta^\kappa=\sum_{\mathrm{cons} \in \mathrm{jet}} \left(\frac{p_{\rm T, cons}}{p_ {\rm T,  jet}}\right)^\kappa r(\mathrm{cons, jet})^\beta ,
\end{equation}
where \ptJet~is the jet's transverse momentum, and $r(\mathrm{cons, jet}) = \sqrt{(\eta_{\rm jet}-\eta_{\rm cons})^2 + (\phi_{\rm jet}-\phi_{\rm cons})^2} $ is the \etaphi distance of a jet-constituent (cons) from the jet-axis (jet). Parameters $\kappa$ and $\beta$ tune experimental sensitivity to hard and wide-angle radiation, respectively. $\lambda^1_{\beta > 0}$ are infra-red and collinear (IRC) safe angularites~\cite{Larkoski_2014}, which probe the average angular spread of energy around the jet-axis.

The jet angularity based observables like jet-substructure measurements in Pb+Pb collisions at \sGevPbPb~at the LHC, have shown that quenched jets, on average, have migration of charged energy away from their axis relative to a \pp baseline \cite{CMS_2014243} and possibly a survivor bias toward harder, quark-like fragmentation \cite{Alice2018}. Similar measurements using jets with lower \ptJet~at RHIC, will help understand jet-medium interactions in a complementary phase-space region to LHC. In these proceedings, jet girth ($g = \lambda_1^1$) and momentum dispersion $(\ptd  =\sqrt{ \lambda_0^2} )$ are measured using data from Au+Au collisions at \sGevAuAu~collected in 2014 using the Solenoidal Tracker At RHIC (STAR) detector system. We also calculate a non-angularity based jet observable \lesub~which gives a measure of the hardest splitting of the jet:
\begin{equation}\label{Eq:LeSubIntro}
	\lesub = \textit{p}_{\mathrm{T, cons}}^{\mathrm{leading}} - \textit{p}_{\mathrm{T, cons}}^{\mathrm{subleading}} .
\end{equation} 
The analysis is also differential in centrality, which is a measure of the transverse overlap between the colliding nuclei, closely related to the impact parameter. Events with lower percentage of centrality (more transverse nuclear overlap) would on average, produce more QGP and hence would be expected to show greater modification compared to a \pp baseline.

\section{Analysis details}\label{sect:AnalDeets}
Charged-particle tracks and neutral energy depositions (towers) are measured using STAR's Time Projection Chamber (TPC) \cite{Anderson_2003} and Barrel Electromagnetic Calorimeter (BEMC) \cite{BEDDO2003725} detectors respectively. Together, they provide full azimuthal coverage with a pseudorapidity acceptance of $|\eta| \leq 1$.  The tracks and towers are clustered into jets using the anti-$k_{\rm T}$ algorithm with a jet resolution parameter $R = 0.4$, implemented using the FastJet library~\cite{Cacciari_2012}. To suppress contributions of fake tracks and combinatorial background (especially in the context of the larger  heavy-ion background), a ``hard-core'' constituent selection  is applied, as was done in previous STAR analyses \cite{PhysRevC.105.044906}, which only allows tracks (towers) with $c\ptTrack (\EtTower) \geq 2~\mathrm{GeV}$ to be clustered into jets. To enhance jet signal, only High-Tower (HT) triggered events, with at least one tower with $\EtTower \geq 5.4$ GeV are considered. After clustering, only jets completely falling within acceptance ($|\eta_{\rm jet}| \leq 0.6$) are kept. Jets with area, $A_{\rm jet} < 0.4$ are rejected to further reduce the fake jet contribution.

\begin{figure}[h]
	\centering
	\includegraphics[width=0.8\linewidth]{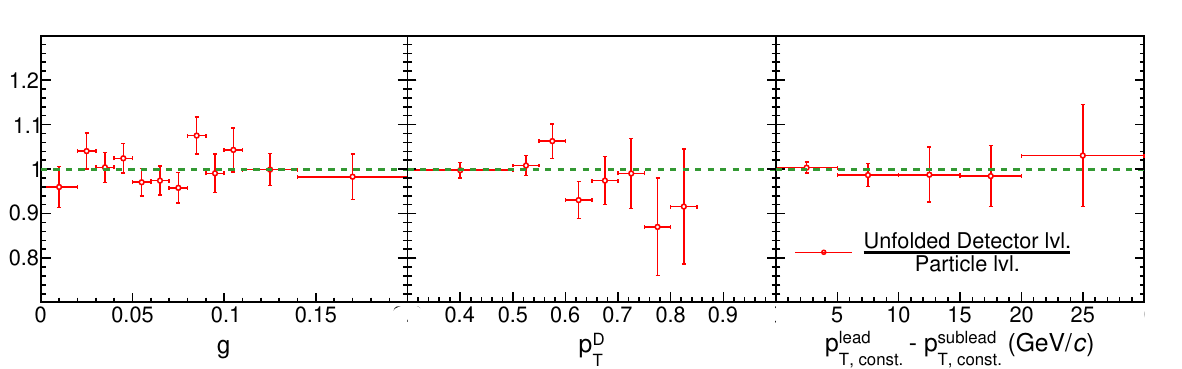}
	\caption{Closure test for \g~(left), \ptd~(middle)~and \lesub~(right) showing the ratio of distributions calculated using jets from unfolded 0-20\% centrality detector level and particle level events from Au+Au collisions at \sGevAuAu.}
	\label{fig:closure}
\end{figure}

Residual background and detector effects are removed by using a mapping between particle and detector level jets from an embedding simulation which involves PYTHIA-6 STAR tune \cite{adkins2019studying} events processed into detector hits using GEANT3 \cite{Brun:1987ma} and added to real minimum-bias events from Au+Au collision environment. MultiFold~\cite{PhysRevLett.124.182001} method was used to simultaneously unfold $p_{\rm T, jet}$, $\eta_{\rm jet}$, $\phi_{\rm jet}$, $N_{\rm cons, charged}$, \g, \ptd~and \lesub. MultiFold uses Dense Neural Networks (DNNs) implemented using the EnergyFlow package~\cite{Komiske_2018},  which are trained on full embedding sample at the detector level and the generator level. MultiFold has been previously shown in~\cite{song2023measurement} to compare well with the more commonly used RooUnfold~\cite{bennerRoounfold}. Closure of the MultiFold is a measure of confidence in the algorithm determined by applying a pre-trained MultiFold on a detector level jet sample for which the particle level is known and comparing the two through ratios. Proper closure is demonstrated in Fig.\ref{fig:closure} for the obervables studied in these proceedings, with the ratios of distributions being consistent with unity within statistical errors.

 This study is the first instance of the MultiFold method being used to deconvolute detector effects from heavy-ion collision events. Efforts are underway to further optimize the algorithm for heavy-ion collisions using hyper-parameter tuning and better feature selection techniques.
 
 \begin{figure}[H]
 	\centering
 	\addOnePanelInFig{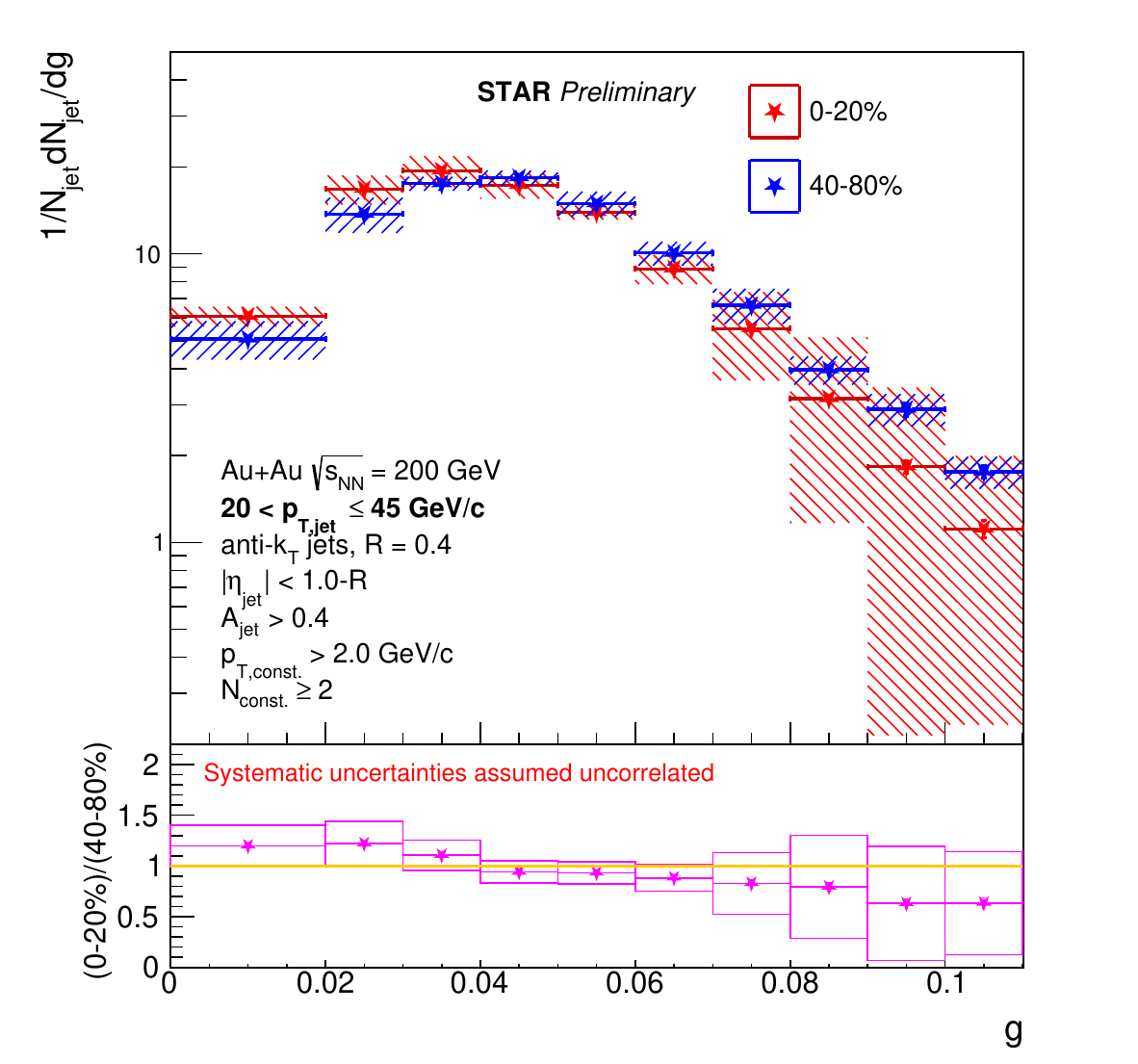}{0.42}
 	\addOnePanelInFig{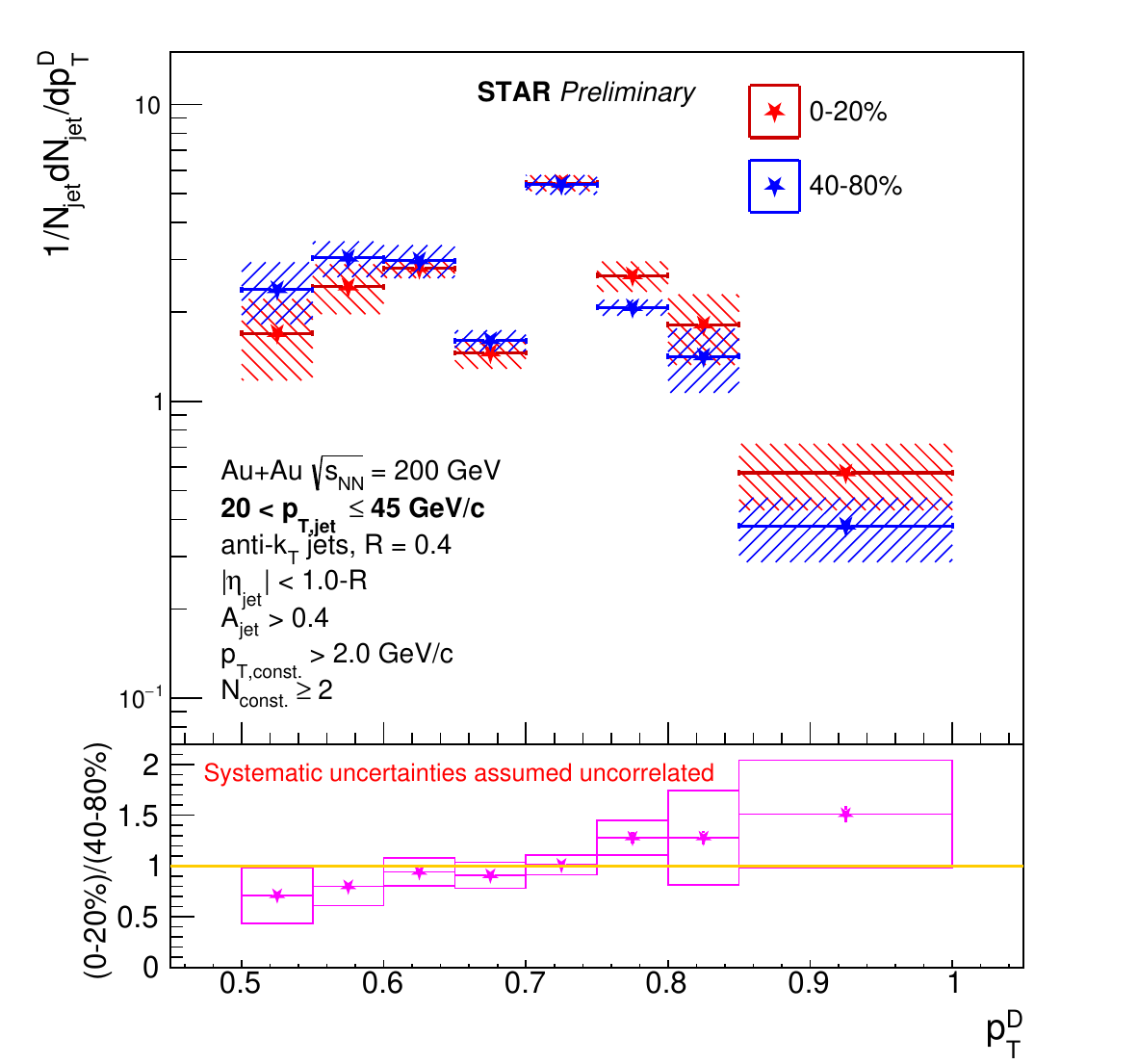}{0.42}
 	\addOnePanelInFig{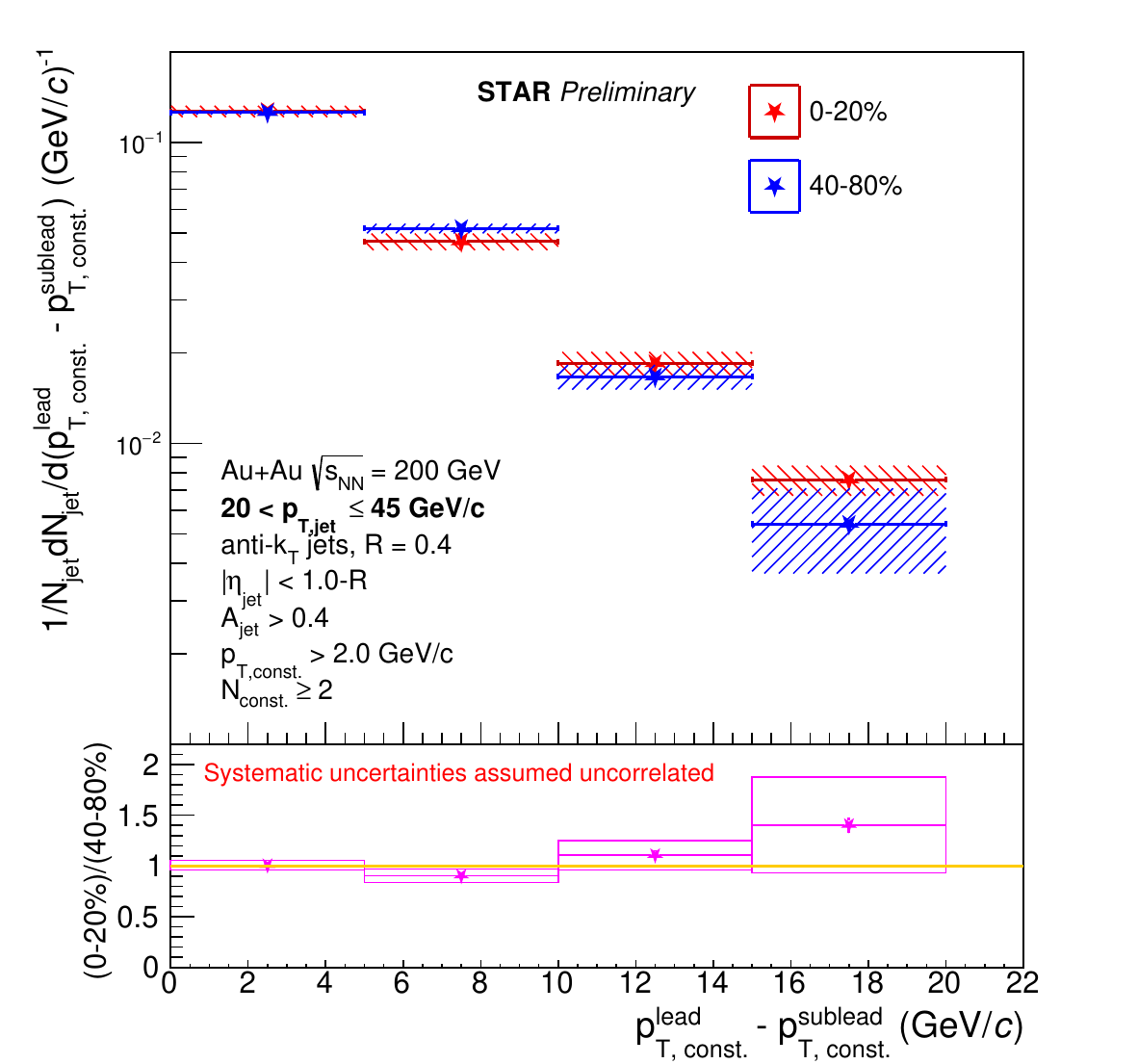}{0.42}
 	\caption{Normalized $g$ (top-left), \ptd~(top-right) and \lesub~(bottom) distributions for jets with \ptJetGeq{20} from central (0-20\%, red stars) and peripheral (40-80\%, blue stars) centrality ranges. Systematic uncertainties are shown as shaded boxes. Central over peripheral ratios (magenta stars) are in the lower panels with their systematic uncertainties represented as magenta boxes. Systematic uncertainties are assumed to be uncorrelated between the centrality ranges. }
 	\label{fig:JSOs}	
 \end{figure}

\section{Result and Discussion}\label{sect:Results}
 Fully corrected measurements of \g, \ptd~and \lesub~distributions calculated from jets with $\ptJet > 20\Gevc$ from events within centrality ranges of 0-20\% (central) and 40-80\% (peripheral), and ratio comparisons between distributions from  central and peripheral events are shown in Fig.~\ref{fig:JSOs}. The \ptd~distribution shows a discontinuity at 0.7 due to the strong dependence on number of constituents in jet. Systematic uncertainties were assumed to be uncorrelated, with uncertainties from various sources being added in quadrature. Parameters varied for calculating the uncertainties in this study are the MultiFold regularization parameters of batch size and number of iterations. Any residual non-closure from MultiFold were also included in the calculation of systematic uncertainties. A prior variation uncertainty was added from previous calculations using STAR $\sqrt{s} = 200$ GeV p+p collision data. 
 
 Results are consistent between central and peripheral collisions within conservative systematic uncertainties. This can be attributed to the assumption of uncorrelated systematic uncertainties, and a bias in the analysis towards hard-fragmented jets due to the selection of events and constituents for jet clustering. Harder-fragmented jets, on average are narrower and have lower interaction time with QGP, which could make any relative modification in central events compared to the peripheral ones lower than what would be potentially observed in an inclusive jet analysis. 

\section{Conclusions} \label{sect:Conclusions}
First fully corrected distributions of $p_T^D$, Girth and LeSub from hard-core jets in heavy-ion collisions at RHIC are presented in these proceedings. With the hard-core jet definition and HT trigger requirement, the sample of jets used here is biased towards hard-fragmented jets. These are also the first heavy-ion results using the multifold technique to remove detector effects and residual background fluctuations. Central and peripheral collisions are compared by taking ratios, which are consistent with unity, within systematic uncertainties. This hints towards low relative quenching of hard-core jets between the central and peripheral events, contrary to the measurements in \cite{Alice2018} from Pb+Pb collisions at  \sGevPbPb~which show significant substructure modifications of small radius jets in QGP.  It must be noted that there is significant room to improve the central over peripheral ratio determination by studying systematic uncertainties in more detail, for which further analyses are ongoing. 


\end{document}